\documentclass[12pt]{article}
\usepackage{a4wide}
\usepackage{bm}
\usepackage{amsfonts}
\def\p{\partial}
\def\a{\alpha}

\def\d{\delta}
\def\e{\varepsilon}

\def\l{\lambda}
\def\r{\rho}
\def\g{\gamma}
\def\o{\omega}

\def\th{\theta}
\def\z{\zeta}

\def\ra{\rightarrow}

\def\Mfunction#1{\mathop{\rm #1}\nolimits}

\begin{document}

\title{Turbulence model of the cosmic structure}
\author{Jos\'e Gaite\\
{\small IDR, ETSI Aeron\'auticos, Universidad Polit\'ecnica de Madrid,}\\ 
{\small Pza.\ Cardenal Cisneros 3, E-28040 Madrid, Spain}
}

\date{January 24, 2012}

\maketitle

\begin{abstract}
The Kolmogorov approach to turbulence is applied to the Burgers turbulence in
the stochastic adhesion model of large-scale structure formation.  As the
perturbative approach to this model is unreliable, here is proposed a new,
non-perturbative approach, based on a suitable formulation of Kolmogorov's
scaling laws.  This approach suggests that the power-law exponent of the
matter density two-point correlation function is in the range 1--1.33, but it
also suggests that the adhesion model neglects important aspects of the
gravitational dynamics.
\end{abstract}

The large-scale structure of the universe is produced by the gravitational
clustering of an initially homogeneous matter distribution.  This process can
be described by the Newtonian equations of motion of the matter fluid, written
in comoving coordinates and in terms of the peculiar velocity and
gravitational fields \cite{Pee}. These equations are nonlinear and, although
they can be linearized to describe the early growth of small perturbations of
the initially homogeneous distribution, the actual structure formation takes
place when the nonlinearity plays a major r\^ole, at the aptly called
nonlinear stage of gravitational clustering.  The nonlinearity of fluid
mechanics also plays a major r\^ole in the phenomenon of turbulence and this
is the cause of the difficulty in treating this phenomenon (often referred to
as the ``unsolved problem of classical physics''). Since turbulence is a key
actor in many astrophysical scenarios, it is tempting to apply methods and
ideas of turbulence to the study of large-scale structure formation.

An early attempt to apply Kolmogorov's scaling laws to the origin of galaxies
was made by Weizs\"acker \cite{Weizs}, but his ideas have been since mostly
restricted to intragalactic turbulence and have played no role in the study of
the formation of clusters and superclusters of galaxies or the large-scale
distribution of the dark matter.  However, a popular model of large-scale
structure formation, namely, the adhesion model \cite{Gurb-Sai,Shan-Zel}, is
essentially a model of pressure-less turbulence, namely, the type of
turbulence that occurs in strongly compressible flows and is usually called
Burgers turbulence.  As shown in this letter, the Kolmogorov approach to
turbulence can be applied to cosmic Burgers turbulence, employing a suitable
formulation of the adhesion model.

In cosmology, scaling laws for the velocity field, such as Kolmogorov's laws,
are especially important if they can be related to scaling laws for the matter
density field, because the positions of astronomical objects are more easily
measurable than their velocities and indeed have been shown to follow scaling
laws. The best known scaling laws in cosmology have been found in the
distribution of galaxies \cite{Syl,Jones-RMP}, but evidences of similar
scaling laws in the dark-matter distribution are found in $N$-body
cosmological simulations.  To be precise, these simulations show that both the
dark matter and the baryonic matter form a common {\em multifractal} ``cosmic
web'' structure \cite{I}.  The cosmic web is precisely the type of structure
predicted by the adhesion model \cite{Gurb-Sai,Shan-Zel}.

The cosmic web consists of sheets (Zeldovich ``pancakes''), filaments and
nodes, so it is indeed multifractal, in the sense that it is formed by objects
of several dimensions, namely, two, one and zero dimensions.  However, such a
distribution needs not be scale invariant. Nevertheless, the actual cosmic web
structure is, arguably, a self-similar multifractal \cite{I}.  The reason why
the adhesion model does not predict a distribution with definite scaling
properties is that the Burgers equation is {\em integrable}, in sharp contrast
with the Navier-Stokes equation of incompressible turbulence. In other words,
the {\em chaotic} properties of the latter are not present in the former, which
nicely evolves the initial conditions. Therefore, the natural way of producing
a self-similar cosmic web structure is to use self-similar initial
conditions, namely, an initial Gaussian velocity distribution with a power-law
power spectrum \cite{V-Frisch}. This type of distribution evolves to a
self-similar cosmic web.

The self-similar cosmic-web solution is obtained from the exact integral of
the Burgers equation in the zero-viscosity limit \cite{V-Frisch}. In this
limit, the Burgers equation is indeed scale invariant, in the sense that
simultaneous space and time scalings $\l x$ and $\l^{1-h} t $ and the induced
scaling $\l^{h} \bm{u}$ leave the equation invariant. Hence, the solution
fulfills the dynamical scaling law
\begin{equation}
\bm{u}(\bm{x}, t ) =  t ^{h/(1-h)} \bm{u}(\bm{x}/ t ^{1/(1-h)},1),
\label{dyn-s1}
\end{equation}
that is, the solution at any $t$ is obtained by scaling the solution at $t=1$.
This scaling is connected with a dynamical invariant, $u^{1/h}/x$, which can
be identified with the specific dissipation rate $\e$ for $h=1/3$ (the
Kolmogorov scaling).  One can further deduce that there is a homogeneity scale
$L( t ) = t ^{1/(1-h)} L(1)$, such that the cosmic web structure at time $ t $
has only formed on scales smaller than $L( t )$ whereas the initial
homogeneous distribution stays on larger scales.  The homogeneity scale $L$ 
plays a similar r\^ole to that of the integral scale in Navier-Stokes turbulence.

The definition of matter density in the adhesion model is not unique, but
there is an ``analytically convenient'' definition \cite{Sai-Woy} that gives
rise to a straightforward relation between velocity and density scaling laws,
because the density field is expressed in terms of the velocity field:
\begin{equation}
\r(\bm{x}) = \r_0 \det\left[\d_{ij} - \p_i u_j(\bm{x})\right],
\label{r-u}
\end{equation}
where $\r_0$ is the constant initial density.  This expression simplifies to
\begin{equation}
\d\r(\bm{x}) = \r(\bm{x}) - \r_0 = -\r_0 \,\p_i u^i(\bm{x})
\label{dr-u}
\end{equation}
in the linear regime, $|\d\r| \ll \r_0$.  In the nonlinear regime, as shock
waves form, $\bm{u}(\bm{x})$ becomes discontinuous and $\p_i u_j \ra -\infty$.
The shock waves are also {\em caustics} \cite{Gurb-Sai,Shan-Zel}, where matter
accumulates and $\r \ra \infty$, according to Eq.~(\ref{r-u}).

Unfortunately, the above-explained approach has obvious shortcomings: (i) the
initial power spectrum is not a power law; (ii) the adhesion model is just a
simplified model of the gravitational dynamics, which actually {\em is}
chaotic; and (iii) the adhesion of matter into caustics is considered as an
inelastic collision but the dissipated energy is not accounted for.  In fact,
the dissipation in the gravitational dynamics is linked to its chaotic nature:
self-gravitating systems tend to virial equilibrium, which is independent of
the initial conditions and implies that entropy grows in the process (this
process is usually called virialization).  Therefore, it is reasonable to
supplement the adhesion model with a ``noise'', which reverts the lost kinetic
energy, on the one hand, and makes the long-time evolution of the velocities
independent of the initial conditions, on the other hand.  This {\em
  stochastic adhesion model} possesses an attractor characterized by a
dynamical scaling that is independent of the initial conditions, unlike the
one defined by Eq.~(\ref{dyn-s1}). The resulting stationary state, in which
the energy injected on scales $>L$ is dissipated at constant rate $\e$ at the
Kolmogorov scale, is analogous to the stationary state of incompressible
turbulence. However, in Burgers turbulence, the dissipation takes place in
caustics and has more spatial variation than in incompressible turbulence,
producing strong {\em intermittency}. Intermittency leads to deviations from
Kolmogorov's scaling for higher order correlation functions. A beautiful
exposition of Kolmogorov's ideas and of intermittency is given by Frisch
\cite{Frisch}.

The stochastic Burgers equation is well studied, since it appears, in terms of
the velocity potential, as an equation for surface growth (the surface's
height is given by the potential). The equation for the potential, called the
KPZ equation, includes a Gaussian noise with power spectrum $D(k,\o)$.  It has
been studied with renormalized perturbation theory \cite{Medina}.  With this
method, the types of noise that give rise to dynamical scaling are determined
as fixed points of the dynamical renormalization group.  For white noise, in
three spatial dimensions, the nontrivial fixed point is repulsive, so the
nonlinear term of the Burgers equation is {\em irrelevant} (in the
renormalization group sense) and the viscous term dominates in the
perturbative stationary state.  Therefore, turbulence can only occur in the
strong-coupling, non-perturbative regime.  The addition of ``colored'' noise
with power-law spectrum $D(k) = D k^{-2\r}$ \cite{Medina} does not improve the
situation: there can be several fixed points, but, in three dimensions, only
the trivial fixed point is stable and only if $\r$ is small; otherwise, it
becomes a saddle point. This means that a noise with sufficient power on large
scales inevitably leads to a strong-coupling stationary state.

The method of Medina et al \cite{Medina} has been adapted to the cosmology
setting by Dom{\'\i}nguez et al \cite{cosmoKPZ}. They consider noise with
power-law spatial and {\em temporal} correlations, $D(k,\o) = D k^{-2\r}
\o^{-2\th}$, and one more coupling (a sort of QFT ``mass'').  The
corresponding renormalization group equations have several fixed points, but
only one is stable.  A choice of $\r$ and $\th$ in certain ranges yields
exponents for the power-law velocity-potential correlation function such that
the corresponding exponents of the density correlation function, obtained
through Eq.~(\ref{dr-u}), fit the range of measured exponents $\g$ of the
two-point correlation function of galaxies.  This intriguing derivation of $\g$ has
several questionable aspects, besides the ad-hoc choice of $\r$ and $\th$.
First, Galilean invariance is broken, as the existence of the stable fixed
point demands a non-zero $\th$ \cite{cosmoKPZ}.  Second, Eq.~(\ref{dr-u}) is
only valid in the linear regime, in principle.  Third, the results, apart from
the values of $\g$, are also questionable: Regarding the values of the
couplings at the fixed-point, the strength $D$ of the correlated part of the
noise (dominant for small $k$) turns out to be negative.  Furthermore, the
``mass'' scale is non-vanishing, so the stable fixed point does not seem to
correspond to a scale invariant stationary state.

At any rate, one can argue, on general grounds, that perturbation theory
(especially, the one-loop approximation) is not the right approach to Burgers
turbulence.  The effective coupling constants in the renormalization group
equations have the generic expression $\l^2D/\nu^3$ (except the ``mass''),
where $\l$ is the nonlinear coupling constant (to be set to the value of
unity), $D$ is a noise strength, and $\nu$ the viscosity. Therefore, as the
nonlinear term dominates in the inertial range, the coupling must be
strong. More precisely, the given expression implies that the coupling
constants are actually proportional to the cube of the Reynolds number, which
has to be a very large number, making perturbation theory unreliable.

Therefore, one must resort to non-perturbative methods.  Standard
non-perturbative methods in turbulence are the closure approaches, in which
the hydrodynamical hierarchy of equations for statistical moments is closed at
some order by assuming a relation between the moments of the corresponding
order and lower order ones.  There is a similar closure approach in cosmology,
based on the BBGKY hierarchy \cite{Pee}. This a second-order closure and it is
consistent with a scaling ansatz for the two-point correlation functions, with
just one power-law exponent, but this number remains undetermined, unless a
connection with an initial power-law power spectrum of perturbations is
assumed.  As we avoid this connection, we prefer to follow the traditional
non-perturbative methods in turbulence. They are based on reasonable
assumptions, the simplest ones being Kolmogorov's universality assumptions,
namely, homogeneity, isotropy, and scaling laws for the moments of
longitudinal velocity increments \cite{Frisch}.  These laws state that
\begin{equation}
\langle \left(\d \bm{u} \cdot \bm{r}/r \right)^n \rangle \propto (\e r)^{n/3},
\label{K}
\end{equation}
where $\d \bm{u} = \bm{u}(\bm{x}+\bm{r}/2) -\bm{u}(\bm{x}-\bm{r}/2)$ and $n
\in \mathbb{N}$.  A general form of these scaling laws, suitable for
introducing the effect of intermittency, is
\begin{equation}
\langle \left|\d \bm{u}\right|^q \rangle = A\, r^{\z(q)}, 
\label{gK}
\end{equation}
where $q \in \mathbb{R}$, and $A$ does not depend on $r$.  The effect of
intermittency is given by the function $\z(q)$ as explained in the following.

Kolmogorov's scaling laws are justified by employing the hierarchy of
hydrodynamical equations, in particular, the second-order one, called the
Karman-Howarth-Monin equation \cite{Frisch}.  A version of this equation is
valid for Burgers turbulence.  An illuminating derivation of the equation has
been given by Polyakov \cite{Polyakov} in the one-dimensional case. Polyakov
realizes that, in the equation for $\p_t{u}^2$, the dissipation in the
inertial range arises as a {\em field-theory anomaly}, due to the
non-differentiability of the velocity field. The form of the anomaly can be
found by employing a point-splitting method. In the three-dimensional case,
the calculation is more involved but it yields the simple result:
\begin{equation}
\p_t  {u}^2(\bm{x}) = -{u}^i(\bm{x}) \frac{\p {u}^2(\bm{x}) }{\p x^i} +
\frac{1}{2} \lim_{\bm{r} \ra \bm{0}} \frac{\p}{\p r^i} \left( \d {u}^i \,\d \bm{u}^2 \right).
\label{Poly}
\end{equation}
The last term is the anomaly $a(\bm{x})$, which would vanish if
$\bm{u}(\bm{x})$ were differentiable.  Remarkably, to derive Eq.~(\ref{Poly}),
we do not need homogeneity or isotropy.  Anyway, homogeneity and isotropy are
part of Kolmogorov's universality assumptions and are natural in cosmology.
From Eq.~(\ref{Poly}), one deduces that the average dissipation in the steady
state is $\e = -\langle a \rangle/2$.  This closure relation is an exact
formulation of the $n=3$ case of the scaling laws (\ref{K}) for Burgers
turbulence, analogous to Kolmogorov's ``$4/5$'' law of incompressible
turbulence \cite{Frisch}. Therefore, in Eq.~(\ref{gK}), $\z(3)=1$ (assuming
that $\e$ is well defined in the limit $\nu \ra 0$).

If the probability $P(\d \bm{u})$ were Gaussian, then $\z(q) \propto q$, and
necessarily $\z(q)= q/3$, as in Eq.~(\ref{K}).  However, intermittency
manifests itself in a slower growth of $\z(q)$ for $q > 3$ \cite{Frisch}.  In
one dimension and with power-law correlated noise, the extent of intermittency
in Burgers turbulence depends on the noise exponent, but it is always so
strong that the maximum of $\z(q)$ is $\z=1$ \cite[especially, Fig.~2]{HJ}.
Generalizing the results of Hayot and Jayaprakash \cite{HJ} to three
dimensions, in terms of the KPZ noise exponent $\r$, the value $\r=5/2$ is
such that the noise strength $D$ has the dimensions of $\e$ and the (Burgers)
noise correlation function is proportional to $\log r$.  This leads to the
Kolmogorov scaling law $\z(q) = q/3$ for $q \leq 3$ \cite{HJ}.  The limit of
the noise correlation function as $r \ra 0$, equal to $\e$, diverges for $\r <
5/2$ and is ill-defined for $\r > 5/2$, in the limits $\nu \ra 0$ or $L \ra
\infty$, respectively.  In other words, $\e = \int_0^\infty k^2 D(k) \,d^3k$
diverges at $k=\infty$ or at $k=0$ and therefore is not universal.  However,
the $r$-dependent part of the noise correlation function is universal and
proportional to $r^{2\r-5}$ if $3/2 < \r < 7/2, \; \r\neq 5/2$.  The values
$5/2 < \r < 7/2$ correspond to large-scale forcing, such that $\z(q)=1$ if $q
\geq 3$, but $2/3<\z(2)<1$.  For $\r > 7/2$, the $r$-dependent part of the
noise correlation function is not universal and depends on scales $> L$; that
is to say, it depends on the initial conditions. Furthermore, $\z(q)=1$ for $q
\geq 2$, in this case.  In this range of $\r$, the stochastic adhesion model
is presumably equivalent to the ordinary adhesion model with self-similar
initial conditions.  Therefore, the interesting range for the cosmic structure
is $\r \in(5/2,7/2)$.  Note that the noise correlation function does not need
to be a power law: higher powers of $r$ are not relevant in the inertial range
and, besides, are not universal.

Our next step is to calculate the density correlation function from
Eq.~(\ref{r-u}), assuming Eq.~(\ref{gK}) with $\z(q)=1$ for $q \geq 3$ and
$2/3<\z(2)<1$.  The expansion of the determinant in Eq.~(\ref{r-u}) yields:
{\setlength\arraycolsep{2pt}
\begin{eqnarray*}
\d\r/\r_0  = 
- \p_i u^i + 
\left(\p_1 u^1 \p_2 u^2 - \p_1 u^2 \p_2 u^1 +  
\p_1 u^1 \p_3 u^3 
{}- \p_1 u^3 \p_3 u^1 + \p_2 u^2 \p_3 u^3 - \p_2 u^3 \p_3 u^2 \right)
\nonumber\\
{}+ \Mfunction{O}(\p u)^3.
\end{eqnarray*}
}%
While Eq.~(\ref{dr-u}), of $\Mfunction{O}(\p u)$, is valid in the linear
regime, we have to consider the formation of caustics.  Caustics are due to
the blowing up of the eigenvalues of the matrix $\p_i u^j = \p_{ij} \phi$,
where $\phi$ is the velocity potential ($\bm{u}$ is discontinuous). If only
one eigenvalue diverges, the collapse is one-dimensional and a sheet forms, so
Eq.~(\ref{dr-u}) is justified. For filaments or nodes, more terms are
necessary.

The reduced two-point correlation function of the density is
{\setlength\arraycolsep{2pt}
\begin{eqnarray}
\langle \d\r(\bm{r}) \, \d\r(\bm{0}) \rangle/\r_0^2  = 
\langle\p_i u^i(\bm{r})\, \p_j u^j(\bm{0}) \rangle - 
2 \,c(r)
{}+ \langle\Mfunction{O}(\p u)^4 \rangle + \cdots,
\label{<rr>}
\end{eqnarray}
}%
where 
{\setlength\arraycolsep{2pt}
\begin{eqnarray*}
c(r) &=& 
\langle\left(\p_1 u^1 \p_2 u^2 - \p_1 u^2 \p_2 u^1 +
\p_1 u^1 \p_3 u^3 - \p_1 u^3 \p_3 u^1 
+ \p_2 u^2 \p_3 u^3 - \p_2 u^3 \p_3 u^2 \right)\!
(\bm{r}) \,\p_j u^j(\bm{0}) \rangle. 
\end{eqnarray*}
}%
We have shown explicitly only terms up to $\Mfunction{O}(\p u)^3$, because the
other terms do not require any calculation, as we now explain.  The functions
on the right-hand side of Eq.~(\ref{<rr>}) are power-laws of $r$, each one
with a characteristic exponent $-\g$ that can be deduced from Eq.~(\ref{gK});
namely, $-\g = \z(n)-n$ for $\langle\Mfunction{O}(\p u)^n \rangle.$ Given that
$\z(n) = 1$ for $n=3,4,5,6,$ we have $\g = 2,3,4,5$, respectively. However,
the maximal value is $\g = 3$, which is the value for a Poisson distribution
(shot-noise) term: this term can appear as either $\d(\bm{r})$ or $r^{-3}$
(see, e.g., \cite{ApJL}). As $\g=3$ is reached for $n \geq 4$, we only need to
consider the cases $n=3$ and $n=2$.

To explicitly calculate $c(r) \propto r^{-2}$, it is useful to express it  as 
$c(r) = \bigtriangleup g(r)$, where 
{\setlength\arraycolsep{2pt}
\begin{eqnarray*}
g(r) &=& 
\langle\left(\p_{11} \phi\, \p_{22} \phi  - \p_{12} \phi\, \p_{21} \phi + 
\p_{11} \phi\, \p_{33} \phi - \p_{13} \phi\, \p_{31} \phi 
+ \p_{22} \phi\, \p_{33}\phi 
- \p_{23} \phi\, \p_{32} \phi\right)\!(\bm{r}) \,\phi(\bm{0}) \rangle. 
\end{eqnarray*}
}%
This function is a dimensionless scalar, so it must be a constant.  Therefore,
$c(r) = 0$ and the $\g = 2$ contribution vanishes.

In the end, the relevant contribution to the density two-point correlation
function is due to the velocity two-point correlation function. The
corresponding exponent is $\g = 2-\z(2)$. Since $2/3 < \z(2) < 1$, we obtain
$1 < \g < 4/3$.  The Kolmogorov scaling $\z(2) = 2/3$ yields the upper bound,
$\g = 4/3 \simeq 1.33.$ The range of values of $\g$ obtained from galaxy
surveys or $N$-body cosmological simulations is (mostly) in the interval
$(1,2)$ \cite{Syl,Jones-RMP,I}. However, the classic value $\g=1.7$, which
still stands \cite{Jones-RMP,I}, is larger than $4/3$.  Nevertheless, a
specific methodology for the analysis of galaxy catalogs \cite{Syl} yields
values of $\g$ in the interval $(1,1.3)$.

To obtain values of $\g$ in the interval $(4/3,2)$, we could take $\r
\in(3/2,5/2)$. Then, the Kolmogorov scale could not be set to zero, so it
should be kept and, preferably, identified with a physical scale.  In the
gravitational dynamics, there is no intrinsic small scale, but there are small
scales in the initial conditions.  In $N$-body cosmological simulations, the
most suitable small scale is the scale of gravitational smoothing. In any
case, if we were to take $\r \in(3/2,5/2)$, then $\z(3) < 1$, so $c(r)$ would
not vanish.  When the density correlation function is the sum of different
powers of $r$, the most singular term dominates in the nonlinear domain $r \ll
L$ (the inertial range).  The most singular term is, of course, the Poisson
term, but it must be discarded \cite{ApJL}.  The next singular component is
$c(r)$, although it vanishes for $\r > 5/2$.  In contrast, for $\r < 5/2$,
$c(r)$ would not vanish and would lead to a $\g > 2$, instead of a $\g
\in(4/3,2)$.

With $1 \leq \g < 4/3$, the density correlation function is dominated by
sheets.  Interestingly, the analysis of $N$-body cosmological simulations
leads to a similar conclusion: the bulk of mass belongs to sheets \cite{I}.
However, to fully understand the r\^ole of the three types of cosmic-web
singularities, namely, sheets, filaments and nodes, one must go beyond the
scope of the adhesion model, because the three types of singularities are very
different in regard to the gravitational dynamics.  The accumulation of matter
in sheets leads to density singularities, but the gravitational potential
stays finite. In contrast, filaments and nodes are gravitational singularities
as well as density singularities, so their formation involves the dissipation
of an infinite amount of energy.  Therefore, it is not surprising that the
analysis of $N$-body cosmological simulations \cite{I} shows that the spectrum
of local dimensions $\a$ is cut off at $\a = 1$, which is precisely the local
dimension of filaments.  For filaments, the gravitational potential has just a
logarithmic singularity, which is milder than the $r^{-1}$ singularity of
nodes and, hence, involves less dissipation.

At any rate, gravitational singularities cannot be described in a Newtonian
framework and need the Theory of General Relativity. In this theory, the
energy dissipated in, for example, the formation of a point singularity as a
black hole is finite, namely, it is given by the well-studied black-hole
entropy. Plausibly, a good part of the gravitational energy dissipation at the
(cosmic) Kolmogorov scale can be attributed to the formation and growth of
super-massive black holes, which occur due to dissipative processes in the
dark matter and, preferentially, in the baryonic matter.  However, the
formation and growth of black holes or other relativistic gravitational
singularities is beyond the scope of the adhesion model and even beyond the
scope of (state-of-the-art) $N$-body cosmological simulations.

In conclusion, a hydrodynamic closure approach to three-dimensional Burgers
turbulence leads to Kolmogorov's scaling laws, although in a general form
compatible with the presence of intermittency.  These scaling laws can be
applied to the stochastic adhesion model of the cosmic structure, in
particular, to the determination of the density two-point correlation
function. The result is in partial agreement with the two-point correlation
function obtained from the distribution of galaxies and from $N$-body
simulations but suggests that the adhesion model underestimates the
contribution of low dimensional singularities (filaments and nodes) to energy
dissipation, whereas $N$-body simulations overestimate it.  It is probably
necessary to have a better modeling of small-scale dissipative processes, and
this modeling may require ingredients from general relativity.

\end{document}